\documentstyle[aasms4]{article}

\begin{document}
\baselineskip=5.5mm

\title{Jets from Time-Dependent Accretion Flows onto a Black Hole}
\author{Koji Nobuta and Tomoyuki Hanawa}
\affil{Department of Astrophysics, School of Science, Nagoya University,
Chikusa-ku, Nagoya 464-8602 Japan}
\authoremail{nobuta@a.phys.nagoya-u.ac.jp and hanawa@a.phys.nagoya-u.ac.jp}
\received{1997 September 10}
\accepted{1998 July 19}

\bigskip
\begin{center}
\large \bf Abstract
\end{center}
We investigate time-dependent
inviscid hydrodynamical accretion flows onto a black hole
using numerical simulations.
We consider the accretion that consists of 
hot tenuous gas with low specific angular momentum and
cold dense gas with high specific angular momentum.
The former accretes continuously and the latter highly intermittently as blobs.
The high specific angular momentum gas blobs
bounce at the centrifugal barrier 
and create shock waves. The low specific angular momentum gas
is heated at the shock fronts and escapes along the rotation axis.
The outgoing gas evolves into pressure-driven jets.
Jet acceleration lasts until the shock waves fade out.
The total amount of the mass ejection is about $ 1-11 $\% of 
the mass of the blobs.
The jet mass increases when the gas blobs are more massive or
have larger specific angular momentum.
We get narrower well-collimated jets when the hot continuous flow
has a lower temperature.
In the numerical simulations we used a finite difference code
based on the total variation diminishing scheme.
It is extended to include the blackbody radiation and
to apply the multi time step scheme for time marching.


\section{INTRODUCTION}

It has been widely believed that an accreting black hole
is the central engine of active galactic nuclei (AGNs).  
The accretion is thought to be the origin of X- and $ \gamma $-ray emission 
and jets emerging from AGNs.  It has been investigated with 
various types of numerical simulations.
Wilson (1972) initiated hydrodynamical simulation of gas accretion
onto a black hole.  His simulation was followed by Hawley, Smarr, \&
Wilson (1984a, 1984b), Clarke, Karpik, \& Henriksen (1985).
Uchida \& Shibata (1985) and Shibata \& Uchida (1986) initiated
magnetohydrodynamical simulations of 
accretion disks threaded by magnetic field.
In their simulations a part of the disk is accelerated by
magneto-centrifugal force and evolves into conical bipolar jets.
Eggum, Coroniti, \& Katz (1988) investigated super-Eddington
accretion flow with taking account of radiative force.
The radiation pressure accelerates jets in their
simulation.

Although some of the previous simulations succeeded in
the jet formation, they have not fully taken account of
the relation between the jets and time variability.
Recent observations suggest a close link between
the jets and time variability.  Some AGNs associated with
jets show intense flares of very short duration.
The BL Lac objects, Mrk 421 and Mrk 501, showed X-ray and $ \gamma
$-ray flares of which duration is only 30 minutes to several hours
(Kerrick et al. 1995 and Macomb et al. 1995 for Mrk 421;
Catanese et al. 1997 for Mrk 501).
Since the $ \gamma $-rays are nonthermal, the
flares are likely to be generated by the shock waves. 
The short duration of flares indicates that the shock waves
generating $ \gamma $-rays are transient and formed
near the black hole.  If a transient shock waves forms near
the black hole, it will affect the mass ejection, i.e., 
jets.  

Since the X- and $ \gamma $-rays
are highly variable on the timescale of hours, the accretion
is also expected to be highly variable.  Although many simulations
have been performed for time-dependent accretion onto a black hole,
most of them study approach to a steady state and the variability 
of the flow is only mild (Molteni, Lanzafame, \& Chakrabarti 1994).
In this paper we investigate highly variable accretion onto a black hole
with numerical simulations.

Using numerical simulations we study accretion of dense
gas clouds onto a black hole.  We assume that the dense
gas clouds are formed by some unknown instability or by
disruption of a larger cloud.  They will spiral during the accretion and 
will be stretched in the azimuthal direction
because of differential rotation.  The stretched gas cloud becomes
a tightly wound spiral similar to a torus around the black hole.
We approximate the stretched 
spiral cloud as a gas torus in our simulations, assuming
the axial symmetry to simplify the models and to save computation time.

Our numerical simulation code has several advantages over
those of the predecessors.  First, we employ the TVD scheme 
(the acronym of Total Variation Diminishing, see, e.g., Hirsch 1988) 
to compute a shock wave without numerical oscillations.
We modified a typical TVD scheme, the Roe (1981) scheme,
for a general equation of state.  Second, our simulations 
cover a wide region from $ r =  2 r_g $ to $ 104 r_g $ with
equal to or more than 800 grid points in the $ r $-direction
and 100 grid points in the $ \theta $-direction, where
$ r_g $ denotes the Schwarzshild radius of the black hole.
The dynamic range is very large in the $ r $-direction and
the spatial resolution is very high ($ \Delta r / r \le 9.5 \times 10^{-3} $
and $ \Delta \theta \le 1.6 \times 10^{-2} $).  
To follow the
time evolution accurately we employed the multi grid scheme
of Chiang, van Leer, \& Powell (1992) that achieves 
the second order accuracy and the conservation of mass,
momentum and energy on a nonuniform numerical grid.
These improvements contribute much to the quality of our 
numerical simulations.  As shown later we find jets emanating 
along the rotation axis
in our numerical simulations.  The jet formation
and the propagation are simulated 
without numerical oscillations and strong numerical dampings.

In \S 2 we describe our models and the methods of computations.
The results of numerical simulations are shown in \S 3.  
We find jets accelerated by shock waves in  most of our numerical 
simulations.  We discuss the formation of jets in \S 4.
Comparisons with preceding numerical simulations and observations
are also given in \S 4.  A Short summary is given in \S 5.
In Appendix we present a Roe (1981) type scheme
for a general equation of state.

\section{MODELS}

We investigate time-dependent accretion flows onto a
black hole having the mass, $ M $.  
For simplicity the gravity of the black hole is 
approximated by the pseudo-Newtonian potential (Paczy\'nski \& Wiita 1980),
\begin{equation}
\Phi ( r ) = - { G M \over r - r_g } \; ,
\label{pseudo-Newtonian}
\end{equation}
where $ G $ denotes the gravitational constant.  
The Schwarzshild radius, $ r_g $, is defined as 
\begin{equation}
r_g = {2 G M \over c^2} \; ,
\end{equation}
where $ c $ is the speed of light. 
This pseudo-Newtonian potential reproduces salient features
of the Schwarzshild black hole.  A particle having a small
specific angular momentum enters into the black hole without
centrifugal bounce both in the Schwarzshild metric and
in the Newtonian dynamics with the pseudo-Newtonian potential.
The critical specific angular momentum is the same and
$ \ell_{\rm cr} $ = 2 $ c r _g $.
With the pseudo-Newtonian potential the hydrodynamical equations are
expressed as
\begin{equation}
{\partial \rho \over \partial t} 
  + \nabla ( \rho \mbox{\boldmath$v$} ) = 0 \; , 
\label{mass-con}
\end{equation}
\begin{equation}
\rho {\partial \mbox{\boldmath$v$} \over \partial t} 
  + \rho ( \mbox{\boldmath$v$} \cdot \nabla ) \mbox{\boldmath$v$} 
  + \nabla P
  = - \rho \nabla \Phi \; ,
\label{momentum-con}
\end{equation}
\begin{equation}
{\partial \over \partial t}
\left[ \rho \left( E + {\mbox{\boldmath$v$}^2 \over 2} \right) \right] + 
\nabla \left[ \rho \mbox{\boldmath$v$} \left(
  H + {\mbox{\boldmath$v$}^2 \over 2} 
  \right) \right]
  = - \rho \mbox{\boldmath$v$} \cdot \nabla \Phi \; ,
\label{energy-con}
\end{equation}
where $ \rho $ is the density, $ P $ is the pressure, 
$ \mbox{\boldmath$v$} $ is the velocity, 
$ E $ is the specific internal energy, and $ H $ is the specific enthalpy.
In the above equations we did not take account of the radiative
and the viscous processes for simplicity.  This simplification may
be validated since we restrict ourselves to the dynamical processes.
We assume that the flow consists of the ideal gas and blackbody
radiation.  Accordingly the pressure, specific internal energy,
specific enthalpy, and specific entropy are expressed as 
a function of the temperature, $ T $, and the density;
\begin{equation}
 P = {\rho k T \over \mu m_{\rm H}} + {a T^4 \over 3} \; ,
\end{equation}
\begin{equation}
 E = {1 \over \gamma-1} {k T \over \mu m_{\rm H}} + {a T^4 \over \rho} \; ,
\end{equation}
\begin{equation}
 H = {\gamma \over \gamma-1} {k T \over \mu m_{\rm H}} 
 + {4 a T^4 \over 3 \rho} \; ,
\end{equation}
and
\begin{equation}
 s = {k \over m_{\rm H} \mu} 
 \left( {1 \over \gamma -1 } \log T - \log \rho \right) +
     {4 a T^3 \over 3 \rho} \; ,
\end{equation}
respectively. Here $ \mu $ is the mean molecular weight, 
$ k $ is the Boltzmann constant, 
$ m_{\rm H} $ is the mass of the hydrogen atom, 
$ a $ is the Stefan-Boltzmann constant,
and $ \gamma $ is the ratio of specific heats.
In numerical simulations we set $ M = 10^7 M_\odot $, $ \mu =0.6 $, and
$ \gamma = 5/3 $.

We solved the hydrodynamical equations 
(\ref{mass-con}) $ - $ (\ref{energy-con}) numerically 
in the spherical polar coordinates, $ (r, \theta, \phi) $.
We assume that the flow is axisymmetric and 
symmetric with respect to the midplane.
The computational domain covers from $ r = 2 r_g $ to
$ 104 r_g $ and $ \theta = 0 $ to $ \pi / 2 $.  
The inner boundary at $ r = 2 r_g $ is set so that 
the gas flows with a supersonic radial velocity 
and accordingly no information can propagate into the computational domain 
from the boundary.
The outer boundary at $ r = 104 r_g $
is a fixed one.

The domain of computation is divided into 8 concentric shells.
In a given shell the grid spacings are uniform 
in the $ r $- and the $ \theta $-directions.
The radial grid spacing, $ \Delta r $,  
is twice larger in a given shell than in the inner adjacent shell.  
Accordingly it is 128 times larger in the outermost shell
than in the innermost shell.
Each shell has 100 grid points in the $ r $- and the $ \theta $-directions
in all the models except models E1 and E2.  
In models E1 and E2 the grid points are twice as large as the others.
The radial grid spacing, $ \Delta r $, is smaller than 
$ \Delta r / r \le 9.5 \times 10^{-3} $
in all the shells.

We applied the multi time step scheme of Chiang et al.\ (1992) 
to integrate the hydrodynamical equations. 
In the multi time step method the time step, $ \Delta t $, is 
set separately for each shell so that it is proportional
to $ \Delta r $ in all the shells.  Accordingly the evolution
in a given period is followed with smaller  
time steps in an outer shell.
This method saves computation time and 
reduces numerical diffusion especially in the outer
shells.   If we had used an equal $ \Delta t $ for time marching,
the numerical diffusion would smear out the shock front in the 
outer region owing to the low CFL (Courant-Friedrichs-Lewy) number thereof. 

To calculate the numerical flux we used the extended Roe (1981) type 
approximate Riemann solver for an arbitrary equation of state, 
while Chiang et al.\ (1992) used the original Roe (1981) Riemann solver.
The details of our approximate Riemann solver are given in Appendix.

Our numerical solutions are second order accurate in
time and space.  The total mass, angular momentum, and
total energy are conserved in the domain of computation.
We confirmed that the specific entropy $ s $, 
specific angular momentum $ \ell $, 
and specific total energy $ E_{\rm tot}
(= H + \mbox{\boldmath $ v $}^2/2 + \Phi) $
of a gas element become stationary along the
stream line with an error less than
$ 4 \times 10^{11} $ erg cm$^3$ g$^{-1}$ K$^{-1}$, 
$ 2 \times 10^{-4} c r_g $, and $ 10^{-4} c^2 $, 
respectively, when the accretion rate is 
kept constant for a long time.
The error of the specific entropy corresponds to that of the temperature 
of 70 K, while typical temperature of the flow in our simulations is
of the order of $ 10^{5} $ K.
These numerical errors come from the finite differencing.
The errors decrease in proportional to the square of the grid spacing, 
$ ( \Delta r )^2 $.

We constructed 18 models having different initial conditions.
In most models one or two dense cool gas tori are superimposed
on a hot steady flow. The gas tori are in the pressure
balance with the surrounding flow at the initial state. 
The hot steady flow is characterized by $ P / \rho $ 
evaluated at the outer boundary.
The models are summarized in Table 1.

\section{RESULTS}

\subsection{Typical Example}

In this subsection we concentrate on model A4, a typical
example of our models.  In model A4, one dense gas torus
accretes towards the black hole, and 
bipolar jets emanate along the rotation axis with
a semi-relativistic velocity.

Figure \ref{a2z} shows the evolution of model A4 by the cross sections
on the meridional plane.  They are denoted in the cylindrical
polar coordinates, $ ( \varpi, z ) $.
Figure \ref{a2z}(a) denotes the initial density distribution 
with colors and 
the velocity by arrows.  The density is denoted also by contours
of which levels are spaced 
with an interval of $ \Delta (\log_{10} \rho) = 0.5 $.
The hot gas accretes from the outer boundary with
a constant accretion rate, $ 0.11 M_\odot $ yr$^{-1}$.
On the outer boundary the density and the pressure decrease exponentially 
with increasing $ \theta $, 
\begin{equation}
 \rho (r_{\rm out}, \theta) = 
 \rho (r_{\rm out}, 0) 
 \exp ( - 3 \theta ) \; ,
\end{equation}
where $ r_{\rm out} $ is the radii of the outer boundary.
The $ r $- and the $ \theta $- components of the velocity 
are set to be $ ( v_r, v_\theta ) = ( -2 \times 10^{-3} c, 0 ) $
on the boundary.
The gas inflowing from the outer boundary has
the specific angular momentum of $ \ell = 1.0 c r_g $ 
in the region of $ 75^\circ \leq \theta \leq 90^\circ $ and 
that of $ \ell = 0 $ in the region of $ 0^\circ \leq \theta < 75^\circ $.
The ratio of the pressure to the density is $ 10^{-3} c^2 $ on the boundary.

At the initial stage ($ t = 0 $ hr), 
one dense gas torus is superimposed with the center at $ r = 20 r_g $ 
on the midplane in a steady ambient gas flow.
The gas torus has the mass of $ 6.0 \times 10^{28} $ g.
The density is 
$ 3.5 \times 10^{-13} $ g cm$^{-3}$ at the center of the torus.
The torus is 3.9 times denser 
than the surrounding ambient gas.
The gas in the torus has large specific angular momentum, 
$ \ell = 2.1 c r_g $, 
while the ambient gas has low one, $ \ell \le 1 c r_g $.
Both the torus and ambient gas have negative specific total energy,
$ E_{\rm tot} = -0.002 c^2 $.

At $ t = 1.9 $ hr, the gas torus infalls and is stretched 
in the $ r $-direction [Figure \ref{a2z}(b)].
The gas torus extends from $ r = 2.5 r_g $ to $ 8.5 r_g $.
The gas torus is compressed in the $ z $-direction.
The centrifugal force is not effective at this stage.

The infall of the gas torus
decelerates owing to the centrifugal force.  At 
$ t $ = 2.0 hr, a part of it moves outwards.  
This is the collision and the bounce of the torus 
with the centrifugal barrier
(i.e., the funnel wall of Molteni et al.\ 1994).
For the gas with $ \ell = 2.1 c r_g $ the barrier
is located at $ r = 2.8 r_g $ on the midplane.
At $ t $ = 2.5 hr, the gas torus is split into two vertically,
and the ambient gas inflows between them.
A shock wave forms at the outer edge of the torus.
At the shock front the gas heats up
from $ 1.0 \times 10^{5} $ K to $ 3.0 \times 10^{5} $ K.
The specific total energy 
becomes positive, $ E_{\rm tot} = 0.08 c^2 $.
The shock wave is an oblique one, and the post-shocked gas flows 
along the shock front towards the $ z $-axis. 

The shock front is an arc in the $ ( \varpi, z ) $ plane.
The arc grows in the curvature radius and angular extension.
At $ t = 3.1 $ hr the arc reaches the $ z $-axis,
and the shock wave crosses themselves at $ (\varpi, z) 
= (0, 5 r_g) $ [Figure \ref{a2z}(c)].
At the crossing point the post-shocked gas concentrates
and has high pressure. 
The high pressure pushes gas to outflow 
along the $ z$-axis [Figure \ref{a2z}(d)]. 
The outflowing gas evolves into bipolar jets.
The post-shocked gas has a large specific total energy,
$ E_{\rm tot} = 0.05 c^2 $ 
at $ (\varpi, z, t) = (0, 10 r_g, 4.2{\rm \ hr}) $ 
[Figure \ref{jet7-L-Etot}(b)].
We define jets as gas components with positive radial velocity and 
specific total energy.

The gas outflowing as jets has 
low specific angular momentum [Figure \ref{jet7-L-Etot}(a)] and 
hence was originally of low density.
The gas having high specific angular momentum of $ \ell 
> 2 c r_g $ does not flow near the $ z $-axis 
because of the centrifugal barrier.
The gas torus works as a shield to prevent low specific
angular momentum gas from accreting onto the black hole.
A shock wave forms by the collision of the low 
specific angular momentum gas with the torus.
The post-shocked gas bypasses the gas torus.
Most of the post-shocked gas rushes onto the black hole
without being sustained by the centrifugal force.
A small fraction of the post-shocked gas is
accelerated again near the axis and evolves into bipolar jets.

At $ t $ = 9.8 and 18.5 hr the head of the jet reaches 
$ z = 48 r_g $ [Figure \ref{a2z}(e)] and 
$ z = 80 r_g $ [Figure \ref{a2z}(f)], respectively. 
The jet is collimated more as it propagates.
The gas velocity is $ v_r = 0.2 c $ 
at the head of the jet at $ t $ = 9.8 hr. 
The gas velocity is almost radial 
($ \vert v_r \vert \gg \vert v_\theta \vert $),
and the sound speed is three times smaller than $ v_r $.
The sound speed decreases owing to the expansion.
The collimation will continue although we stopped this
simulation at $ t $ = 18.5 hr. 

Figure \ref{jet7-rho-v} shows the structures on the $ z $-axis at
$ t $ = 9.8 hr.  The radial velocity has two peaks
at $ z = 20 r_g $ and $ 48 r_g $.  The former peak is
the shock front propagating outward and the latter
is that of the rarefaction wave.  The footpoint of
the jet $ ( v_r = 0) $ moves gradually outward from $ z = 7 r_g $
(at $ t $ = 4.8 hr) to $ z = 18 r_g $ (at $ t $ = 9.8 hr).
The evolution of the jets mass is shown in Figure \ref{jet-mass-L1}.
The jet mass increases with the time at a constant
rate, $ 1.8 \times 10^{23} $ g s$^{-1}$.
The constant ejection of the jets starts at $ t = 3.5 $ hr and 
lasts $ 3.2 $ hr.
The mass of the jets decreases gradually in the period of $ t \geq 6.7 $ hr.
The jet energy, which is defined as the positive energy 
$ ( = H + \mbox{\boldmath $ v $}^2/2 ) $ of the jets,
evolves almost proportionally to the jet mass through the simulation.

\subsection{Effects of the Specific Angular Momentum of the Torus}

We constructed models A1, A2, A3, A5, and A6 to study the effects of 
the specific angular momentum of the torus.  
The specific angular momentum is $ \ell $ = 1.9, 1.98, 2.0, 2.2, 
and $ 2.3 c r_g $ in models 
A1, A2, A3, A5, and A6, respectively, 
while it is $ \ell = 2.1 c r_g $ in model A4.
The other model parameters are
the same as those in model A4.

These models can be classified by whether $ \ell $ is lower or higher 
than the critical one, $ \ell_{\rm mb} $.
Here $ \ell_{\rm mb} $ is the specific angular momentum of 
the marginally bound stable orbit (Abramowicz, Jaroszy\'nski, \& Sikora 1978),
and is equal to $ 2 c r_g $ in our model.

When the specific angular momentum of the torus is low (models A1 and A2),
the gas torus accretes 
onto the black hole without experiencing the centrifugal barrier.
It stretches in the $ \varpi $-direction and 
is compressed in the $ z $-direction during the infall.
At $ t $ = 0.93 hr the torus extends from $ \varpi = 6 r_g $ to $ 17 r_g $ 
and is compressed within $ z < 2.5 r_g $ in models A1 and A2. 
At $ t $ =1.4 hr the inner edge of the torus reaches the inner
boundary, $ r = 2 r_g $ in models A1 and A2. 
The radial velocity of the torus is $ 0.5 c $ in model A1
and $ 0.6 c $ in model A2 at the inner boundary 
$ ( \varpi, z ) = ( 2 r_g, 0) $.
The velocities are slower than the radial velocity of the ambient gas, 
$ 0.85 c $, but are still highly supersonic.
By $ t $ = 2.3 hr the gas torus has accreted onto the black hole 
in model A1, and by $ t $ = 2.4 hr in model A2.
The flow becomes steady thenafter.
Shock waves and jets are not formed in models A1 and A2.

When the specific angular momentum of the torus is high (models A3, A5 and A6),
shock waves and jets are formed as well as in model A4.
At $ t = 1.9 $ hr the torus extends 
from $ \varpi = 2.5 r_g $ to $ 8.5 r_g $ 
and is compressed within $ |z| < 2.5 r_g $ in all the models.
A part of the torus collides with a centrifugal barrier 
and moves outwards at $ t = 2.05 $, $ 2.00 $, $ 2.00 $, and $ 1.95 $ hr 
in models A3, A4, A5, and A6, respectively.
At the outer edge of the torus, an arc shock wave forms.
The arc reaches the $ z $-axis, and the shock waves cross themselves at
$ ( \varpi, z ) = ( 0, 5 r_g ) $ in all the models 
and at $ t = 3.15 $, $ 3.1 $, $ 3.1 $, and $ 3.05 $ hr
in models A3, A4, A5, and A6, respectively.
At the crossing point the post-shocked gas has high pressure and 
the pressure driven jets emanate.
At $ t = 7.6 $ hr the head of the jet reaches 
$ z = 32 $, $ 34 $, $ 38 $, and $ 29 r_g $ 
in models A3, A4, A5, and A6, respectively.
At the same time the gas velocities of the jet are
$ v_r = 0.20 $, $ 0.25 $, $ 0.28 $, and $ 0.20 c $
in models A3, A4, A5, and A6, respectively.

Figure \ref{jet-mass-L1} shows the evolution of the jet mass for
models A3, A4, A5, and A6.
The jet ejection starts at the same time of $ t = $ 3.5 hr 
in all the models. 
In model A5 the mass ejection rate is $ 3.0 \times 10^{23} $ g s$^{-1}$ and 
is 1.3 times larger than that in models A3, A4, and A6.
The increase of the jet mass lasts for 2.5 hr in model A3 and 
for about 3.5 hr in models A4, A5, and A6.
Accordingly the resultant jet mass is largest in model A5.
The jet mass gradually decrease in the period of $ t > 6 $ hr in model A3, 
and in the period of $ t > 7 $ hr in models A4, A5, and A6.

\subsection{Effects of the Torus Mass on Jets}

We constructed models B1 and B2 to study the effects
of the torus mass on the jets.  These models are the
same as model A4 except for the density in the tori.  The torus mass 
is $ 3.0 \times 10^{28} $ g in model B1 
and $ 1.2 \times 10^{29} $ g in model B2 while
it is $ 6.0 \times 10^{28} $ g in model A4.

Early stages of models B1 and B2 are similar to those of
model A4.  At $ t $ = 2.0 hr the gas torus collides with
the centrifugal barrier and a part of the gas torus moves outwards 
in all the models. A shock wave forming at the outer edge of
the torus has a larger phase velocity in model B2 and
a smaller one in model B1 than in model A4.
At $ t= $ 3.5 hr the shock wave 
created at the outer edge of the torus reach 
$ \varpi = 8.5 r_g $, $ 11.5 r_g $, and
$ 12.5 r_g $ in model B1, A4, and B2, respectively.
The shock is stronger in model B2
and weaker in model B1 than in model A4.
The maximum value of the specific total energy 
in the post-shocked gas is $ 0.02 c^2 $, $ 0.04 c^2 $, and $ 0.1 c^2 $ 
in model B1, A4, and B2, respectively.
The shock waves cross on the $ z $-axis, and jets form
at $ t $ = 3.5 hr in model B1 
and at $ t $ = 3.1 hr in model B2.  The shock waves
cross at 
$ (\varpi, z) \simeq (0, 5 r_g) $ in all the models.

In model B1 the outflowing gas is decelerated by colliding with 
the inflowing hot gas.  
At $ t $ = 6.3 hr the head of the outflowing gas reaches $ z = 10 r_g $.
Thenafter a part of the gas becomes inflow towards the black hole.
At $ t $ = 6.7 hr the radial velocity is negative 
($ v_r < $ 0) everywhere in the computational domain.
By $ t $ = 7.7 hr most of the hot gas components that were the outflow 
have accreted onto the black hole. 

In model B2 the bipolar outflows evolve into
jets. At $ t $ = 6.9 hr the jets reach $ 50 r_g $. 
The gas velocity is $ v_r = 0.37 c $ and is larger than
that in model A4.  The radial velocity of the jets has two-peaks 
as in model A4.

Figure \ref{jet-mass-m} (a) shows the evolution of the mass of the jets 
in model B1, A4, and B2.
When the torus mass is larger, the mass of the jet increases at a
higher rate and the jet ejection lasts longer.  
The rate of the mass increase is proportional to 
the torus mass as shown in Figure \ref{jet-mass-m} (b).  In model B1
the jet mass is very small and the jets has a very short
lifetime.  There should be a critical mass of $ \sim 10^{28} $ g 
for the torus to induce jets through the shock wave formation.

The jet ejection lasts for $ 4 $ hr and $ 6.5 $ hr 
in models A4 and B2, respectively.
Thenafter the jet mass decreases owing to
the collision with the ambient inflow.
At the final stage of the simulation, $ t > 12 $ hr, the jets begin to 
grow in mass again in model B2.
The increase of the mass is due to snowplow of the ambient gas 
by the strong jets.
The gas torus stays around
$ ( \varpi, z ) = ( 5 r_g, 0 ) $ at the final stages of the simulation.  

\subsection{Accretion of Two Tori}

We constructed models C1, C2, C3, and C4, in which two dense gas tori are 
imposed at the initial stage.  
The inner gas torus is located at $ (\varpi,z)=(20 r_g, 0)$ and 
the outer torus is at $ (\varpi,z)=(30 r_g, 0)$.
Each gas torus has the mass of $ 6 \times 10^{28} $ g.
The inner gas torus has the specific angular momentum of 
$ 2.1 c r_g $ in all the models.
The outer torus has the specific angular momentum of 
1.0, 1.7, 1.9, and $ 2.1 c r_g $ 
in models C1, C2, C3, and C4, respectively.
The other model parameters are the same as those in model A4.

Figure \ref{a2z-c4} shows the evolution of model C4.
All the models look similar with each other
by $ t $ = 3.8 hr, i.e., until 
the inner torus has collided with the outer one
at $ ( \varpi, z ) = ( 10 r_g, 0 ) $.  At that time
the head of the jet reaches $ z = 11 r_g $.
Model C2 is very similar to model C1 even at later stages up to the
ends of the simulations.  The jet mass is only 
5\% different in these models.  The velocity distribution
of the jets is almost the same in these models.

The collision of the outer torus with the inner one 
strengthens the shock wave appreciably in models C3 and C4.
The shock wave forming at the outer edge of the torus 
has a larger phase velocity in models C3 and C4 than that of models C1 and C2.
At $ t $ = 4 hr the outer edge of the 
shock reaches $ r = 16 r_g $ and $ 21 r_g $. 
The specific total energy of the post-shocked is larger 
($ 0.02 c^2 $ in model C3 and $ 0.05 c^2 $ in model C4) than
in models C1 and C2.  After the collision with the outer torus,
the first torus splits into two in the vertical direction.
Each of the two leaves the midplane moving outwards in the
radial direction.

The outer torus forms another shock wave after its collision with
the centrifugal barrier in models C3 and C4.  
This shock wave crosses at $ 5 r_g $ on
the $ z $-axis and creates second jets at
$ ( \varpi, z, t ) = ( 0, 8r_g, 6.4 {\rm \ hr} ) $.
At $ t $ = 7.5 hr the radial velocity of the jet has four peaks 
at $ 9, 16, 27, 42 r_g $ in model C3 and 
at $ 10, 17, 27, 42 r_g $ in model C4.
The outer two peaks are a pair of the shock and
rarefaction waves due to the inner torus, 
while the inner two peaks are those due to the outer torus.
The second jets have a higher out going velocity and overtake
the first jets. 
The second jet is stronger in model C4 than that in model C3.
The gas velocity of the second jet is $ 0.20 c $ in model C3 
and $ 0.35 c $ in model C4.  
When the outer torus has higher specific
angular momentum, $ \ell \ge \ell_{\rm mb} $, 
it strengthens the jets by the formation of the second shock wave.

Figure \ref{jet-mass-L} shows the mass of the jets 
in models C1, C2, C3, and C4 as a function of $ t $.
By $ t $ = 6.4 hr the mass of the jets increases
at a constant rate in all the models.
In model C4 the rate increases by a factor of 2
in the period of 6.4 hr $ < t < $ 10.9 hr.
This enhancement is ascribed to the infall of the outer
torus of high specific angular momentum.
In model C3 the rate increases slightly 
in the period of 6.4 hr $ < t < $ 7.8 hr.
In models C1 and C2, the evolution of the jet mass is almost same
and is also similar to that in model A4.
This implies that the gas blob with low specific angular momentum,
$ \ell < \ell_{\rm mb} $, does not increase the jet mass.

The shock waves that have propagated outward begin to propagate inward 
around $ t \simeq $ 6 hr in models C1 and C2, 
$ t \simeq $ 8 hr in models C3, and 
$ t \simeq $ 11 hr in model C4. 
Thenafter the jet mass remains constant.  
The slight decrease of the jet mass is due to
the deceleration by the interaction with the ambient inflow.
Another increase of the mass in model C4 in the period of $ t > 14 $ hr 
is due to snowplow of the ambient gas 
by the strong jets as well as in model B2.

\subsection{Effects of the Shape of the Tori}

We constructed models D1 and D2 to study the effects of the 
shape of the gas tori.
Models D1 and D2 are the same as model C4 except for the shape 
of the gas torus at $ t = 0 $ hr. 
In model C4 each gas torus has a circular cross section of which
radius is $ 4 r_g $ in the $ ( \varpi, z) $ plane.
In models D1 and D2 each gas torus has a rectangular cross section
of which width is $ 6 r_g $.  Each torus has a half thickness
of $ 8 r_g $ and $ 12 r_g $ in the $ z $-direction in models D1
and D2, respectively.  The mass of the each torus is the same,
and also the pressures in the torus are common 
in models C4, D1, and D2.  The torus density is lowest in model
D2 and lower in model D1 than in model C4.

While the gas tori move inward, they are stretched in the
$ \varpi $-direction and compressed a little in the $ z $-direction
also in models D1 and D2.
The inner gas torus collides with the centrifugal barrier 
at $ t $ = 2.0 hr and returns outward also in models D1 and D2.
A shock wave forms at the outer edge of the inner torus, and
the shock front has a larger vertical extent in models D1 and
D2 than in model C4.

The inner torus collides with the outer torus at $ t $ = 2.5 hr
in models D1 and D2.  The collision has the largest cross section
in model D2 and a larger one in model D1 than in model C4.
The inner torus is decelerated more by the collision in
models D1 and D2 than in model C4.
Also the shock front expands relatively slowly in models D1 and D2.
In models D1 and D2 the shock waves have a larger extent in the $ z $-direction
than that in model C4.
This is because the gas tori have a larger height in models D1 and D2
than in model C4.
Jets start to emerge in models D1 and D2 at $ t $ = 4.5 hr, 
i.e., 1.0 hour later than in model C4.
Figure \ref{jet-mass-spc} shows the evolution of the jet mass
for models D1, D2, and C4.
At the early stages of the jet ejection, 3 hr $ < t < $ 10 hr, 
the mass ejection rates are larger in  models D1 and D2 
by a factor of 1.5 than that in model C4.
In all the models the mass ejection rate increases
slightly at $ t $ = 7 hr, when the second shock wave
formed by the initially outer gas torus reaches the $ z $-axis. 
The mass of the jets reaches the peak at $ t $ = 10 hr in models D1 and D2.

After the collision at $ t $ = 2.5 hr, the initially inner torus
goes outward and the initially outer torus goes inward.
The initially outer torus split two in the vertical direction
and merge again after the collision.  
It collides with the centrifugal barrier and bounces at $ t $ = 4.3 hr. 
After the bounce, 
the initially outer torus fragments, and a small fraction of it
departs from the midplane.  
The initially inner torus  turns to infall again at $ t $ = 9 hr.
The main bodies of the gas tori merge and form
a vibrating compound torus.

Third jets emerge at $ t $ = 12 hr in model D1 and at $ t $ = 12.5 hr
in model D2.  The third jets are initiated by the shock waves
formed by the vibration of the compound torus.
The shock waves cross  
at $ ( \varpi, z, t ) = ( 0, 8 r_g, 12 {\rm \ hr} ) $ in model D1
and at $ ( 0, 8 r_g, 12.5 {\rm \ hr} ) $ in model D2.
The third jets increase the mass of the jets again.
The head of the jet reaches $ 88 r_g $ in model D1 and $ 78 r_g $ 
in model D2 at $ t $ = 15 hr.
At the last stage of the calculation, $ t \approx 17 $ hr, 
the difference of the mass of the jets among three models 
is within a factor of $ 1.5 $. 
The larger mass ejection is due to the large shock wave.

\subsection{Effects of the Sound Speed of the Ambient Flow}

We constructed models E1 and E2 to 
study the effects of the sound speed of the ambient gas.
Models E1 and E2 are the same as models C2 and C4, respectively, 
except for the temperature of the ambient gas.
In models E1 and E2, $ P / \rho $
is $ 5 \times 10^{-4} c^2 $ on the boundary, while 
in models C2 and C4 it is $ 10^{-3} c^2 $.
In models E1 and E2 the flow is concentrated more near the
midplane than in models C2 and C4. 
On the outer boundary the density and the pressure decrease exponentially 
with increasing $ \theta $, 
\begin{equation}
 \rho (r_{\rm out}, \theta) = 
 \rho (r_{\rm out}, 0) 
 \exp ( - 4.2 \theta ) \; ,
\end{equation}
in models E1 and E2.
The accretion rate is
fixed to be the same on the outer boundary in these four models.
At $ (\varpi, z) = (40 r_g, 0 ) $ the
pressure is 35\% lower in models E1 and E2 than in
models C2 and C4, while the density thereof is 40\% higher in
models E1 and E2 than in models C2 and C4.

Figure \ref{a2z-e2} shows the evolution of model E2.
At $ t = 2.0 $ hr the inner gas torus collides with the 
centrifugal barrier and bounces in models E1 and E2.
The total pressure, $ P + \rho \mbox{\boldmath$v$}^2 $, of the ambient gas 
is 1.3 times stronger in models E1 and E2 than 
that in models C2 and C4 at $ ( \varpi, z ) = ( 10 r_g, 0 ) $. 
The high total pressure works against the torus outward
motion and the shock wave propagation.
The shock front reaches only $ \varpi = 13 r_g $
and $ 16 r_g $ at its maximal expansion in models E1
and E2, respectively,  while it reaches
$ \varpi = 15 r_g $ and $ 21 r_g $ in models C2 and C4.
Jets start to emerge at $ t $ = 3.0 hr in these four models from
the point $ (\varpi, z ) = (0, 5 r_g ) $
in which the shock waves cross.
Jets have a higher gas velocity in models E1 and E2
than in models C2 and C4.  At an early stage of
$ t $ = 4.5 hr the outgoing gas velocity is as large
as $ v_r = 0.4 c $ in model E1 and $ 0.5 c $ in
model E2. 
When the ambient gas has a lower sound speed,
the jets have a larger velocity.
At $ t $ = 10.3 hr the head of the jet reaches $ 100 r_g $ 
in model E1 and $ 102 r_g $ in model E2.
Then the maximum velocity is $ 0.40 c $ in model E1 and
$ 0.45 c $ in model E2 (Figure \ref{jet15-rho-v}).
Jets are collimated more in models E1 and E2 than in
models C2 and C4.  The better collimation is mainly due to
the larger Mach number of the jet. 
If the ambient gas were much colder, 
the Mach number of the resultant jet would be much larger 
and thus better collimated jets would be formed.
It should be noted that a higher spatial resolution is required 
to simulate a colder accretion flow.
The grid should be spaced so that 
$ \partial \Phi / \partial r \cdot \Delta r \ll P / \rho $.

Figure \ref{jet-mass-amb} shows the evolution of the jet mass 
for models E1, E2, C2, and C4.
The mass ejection rate is slightly larger and 
the duration of the jets ejection is slightly shorter 
in models E1 and E2 than 
models C2 and C4, respectively. 
Thus, the mass of the jets is larger in models E1 and E2.

\subsection{Sequential Accretion of Many Tori}

We constructed models F1 and F2 by changing
the number of the gas tori.
In model F1, gas tori are sequentially
submitted from the outer boundary with an interval of 2.3 hr.
Each gas torus has the mass of $ 6 \times 10^{28} $ g.
The mass supply by the gas tori amounts 
$ 0.11 M_\odot $ yr$^{-1} $ on average.
In model F2, the gas having the specific angular momentum of $ 2.1 c r_g $ 
inflows continuously from the outer boundary.
At the initial stage of model F2 the density, temperature, 
and radial velocity are 
the same as those of a steady flow except that 
the specific angular momentum is
$ 2.1 c r_g $ in the region of $ r \geq 20 r_g $ 
and $ 75^\circ \leq \theta \leq 90^\circ $.
In model F2 the gas with low specific angular momentum accretes 
at the rate of $ 4.3 \times 10^{-2} M_\odot $ yr$^{-1} $ 
and that with a large one at the rate of 
$ 6.7 \times 10^{-2} M_\odot $ yr$^{-1} $.

In model F1 the jets are formed intermittently. 
At $ t $ = 3.5 hr the first gas torus reaches 
the centrifugal barrier and bounces. The gas torus forms shock waves 
by colliding with the ambient gas. The shock waves 
cross at $ ( \varpi, z, t ) = ( 0, 6 r_g, 5.6 {\rm \ hr} ) $, and 
the jets are formed.
The second and the third gas tori trace the evolution of the first one.
They are stretched in the $ \varpi $-direction during the infall and 
bounce at the centrifugal barrier.
The shock waves formed by the second and the third gas tori
cross at $ t $ = 8.8 hr and 11.8 hr, respectively, and form jets.
At $ t $ = 11.8 hr, the three gas tori coalesce with each other and transform
vertically separated two dense tori.
They reside in the region of 
$ 5 r_g < \varpi < 15 r_g $ and $ 2 r_g < z < 15 r_g $.

At $ t $ = 12 hr the fourth torus collides with the centrifugal barrier 
and bounces.
The fourth torus not only forms shock waves at its head 
but also pushes the coalesced gas upward.
The shock waves by the fourth gas torus do not cross on the $ z $-axis.
The fifth and the sixth gas tori collide 
with the barrier at $ t $ = 13.6 hr and 16.0 hr,
respectively. They also push the coalesced gas upward.
At $ t $ = 17 hr a part of the coalesced gas is expelled along the funnel wall.
The shock waves formed by the fifth and the sixth gas tori cross 
on the $ z $-axis at $ t $ = 14.8 hr and 17.7 hr, respectively, 
and form the jets.

At $ t $ = 18 hr the head of the jet reaches $ z = 93 r_g $.
The gas velocity of the jet is $ 0.20 c $.
The expelled gas blob of large specific angular momentum 
spreads in the region of $ 12 r_g < \varpi < 22 r_g $ 
and $ 20 r_g < z < 35 r_g $.
Its radial gas velocity is $ 0.15 c $.

Figure \ref{jet-mass-spc} shows the mass of the jets
as a function of $ t $ for model F1.
The mass ejection rate is $ 5 \times 10^{23} $ g s $^{-1}$
on average in the period of 7 hr $ < t < $ 18 hr. 
The evolution is similar in models F1 and C3 in the period of $ t < $ 11 hr.
The mass ejection rate increases at $ t $ = 8.8 hr and 11.8 hr, 
when the second and the third gas tori form the jets to emanate.
From $ t $ = 12 hr to 15 hr the ejection rate is smaller than the before.
In this period the arc shock waves do not cross on the $ z $-axis,
and new components of jets are not formed.
At about $ t $ = 15 hr the ejection rate increases again.
This increase is in part due to the ejection of 
large specific angular momentum gas.

In model F2 jets are formed but weak.
At $ t $ = 2.3 hr the head of the high specific angular momentum gas reaches
the centrifugal barrier and bounces. 
The bounce of the gas is much weaker than 
that in model F1. 
The large specific angular momentum gas piles up near the barrier and
forms high density fat torus. 
The inner edge of the torus is bounded by the centrifugal barrier.
The torus is in a quasi-stationary equilibrium.
The ambient gas collides with the torus, 
and the arc shock waves are formed at the interface.

At $ t $ = 3.8 hr the torus occupies the region of 
$ 4 r_g < \varpi < 8 r_g $ and $ |z| < 3.5 r_g $.
The density of the torus is $ 8 \times 10^{-13} $ g cm$^{-3}$ 
and is denser by a factor of 9 than the ambient gas on average.
The specific total energy of the ambient gas increases 
from $ -0.002 c^2 $ to $ 0.005 c^2 $ by passing through the shock front.
The energy gain is smaller than that in model F1.
At $ t $ = 6 hr the torus occupies the larger region of 
$ 4 r_g < \varpi < 12 r_g $ and $ z < 6 r_g $.
The arc shock waves approach to the $ z $-axis as the torus grows.

The arc shock waves cross 
at $ ( \varpi, z, t ) = ( 0, 5 r_g, 6 {\rm \ hr}) $,
and the jets are formed.
At $ t $ = 10.8 hr the head of the jet reaches $ z= 22 r_g $.
The gas velocity of the jet is $ 0.08 c $ and is slower than that in model F1.
The jets are not collimated; the radial width is 
as large as 90\% of the height.
The specific total energy in the jet is $ 0.01 c^2 $ and is smaller than 
that in model F1.

The mass evolution of the jets is shown in Figure \ref{jet-mass-spc}.
The mass ejection rate of the jets increases monotonically.
It is less than that in model F1.

\section{DISCUSSION}

In this section we describe the essence of our model, and
discuss acceleration mechanism in our model,
comparison of our model with other hydrodynamical models,
and application of our model to AGNs and Galactic black hole candidates.

\subsection{Essence of Our Model}

In \S 3 we studied formation of shock wave and jet in highly variable flows
with various numerical simulations.
Although each simulation has different features,
the flow pattern and its dynamics are qualitatively similar.
Figure \ref{jet-acceleration} illustrates the mechanism of
jet formation schematically.  Each panel of Figure
\ref{jet-acceleration} is based on the density distributions in
model A4 at a given epoch.  Panel (a) shows the initial stage 
of model A4.  The colors denote the density distribution.
The dotted curve encircles high specific angular momentum gas.
The dashed curve denotes the centrifugal barrier
for the gas having $ \ell \, = \, 2.1 \, c r _g $ and 
$ E _{\rm tot} \, = \, 0 $.  Since the location of the centrifugal
barrier depends little on $ E _{\rm tot} $, the dashed curve
can be regarded as the centrifugal barrier for the torus gas
($ \ell \, = \, 2.1 \, c r _g $).
Panel (b) denotes the bounce of the torus 
gas.  The colors denote the density distribution at 
$ t $ = 2.9 hr.  The straight and the turning arrows denote
the flow of low and high specific angular momentum gases,
respectively.  The high specific angular momentum gas
collide with the centrifugal barrier (the dashed curve)
and bounces.  After the
bounce it collide with the low specific angular momentum
gas and a shock wave
forms at the collision interface.  The shock front is denoted
with the thick solid curve in panel (b).
Panel (c) denotes the jet formation near the axis.  The colors
denote the density distribution at $ t $ = 6.0 hr.
The shock waves intersects itself on the axis and the shock
front (solid curves) has a complicated structure.  
The radial flow converging to the axis turns out to be
outflow evolving to jets and inflow absorbed to the black hole.
The outflow forms another shock wave traveling in the $ z $-direction.

It is essential for jet formation that the flow contains two
components, i.e., high and low specific angular momentum gases.
The jets are accelerated by shock waves and the shock waves are
formed by the collision of the high and the low specific angular
momentum gases.  The high specific angular momentum gas 
($ \ell \ge 2 c r_g $) bounces at the centrifugal
barrier.  On the other hand the low specific angular momentum gas
($ \ell < 2 c r_g $) 
tends to accrete onto the black hole without centrifugal bounce.
Eventually they collide with each other to form a shock wave.
By the collision the high specific angular momentum gas looses
its kinetic energy while the low specific angular momentum 
gas receives it.  This collision is the first step of the
jet formation in our model.  The second step is the 
intersection of the shock wave on the axis.  When the shock
wave intersects itself, a part of the low specific angular
momentum gas gains energy from the rest and evolves
into the jets.  In the next subsection we evaluate the
energy gain at each step.

\subsection{Acceleration and Energy Gain}

As shown in the previous section, high energy gas is
ejected as jets in our numerical simulations.
The jets consist of gas that had low energy and
was gravitationally bound at the initial state.
The jets are accelerated hydrodynamically and 
gain energy from gas accreting onto the black hole.
In this subsection we discuss how the jet gas gains energy.
We also evaluate the efficiency of the energy gain.

   As shown in Figure \ref{jet-acceleration}
the jet gas is accelerated twice at the encounters with
the shock waves.  The jet gas was originally the ambient
gas flowing near the midplane.
First the ambient gas passes through the shock wave at the collision
with the gas torus bounced at the centrifugal barrier.
Later it passes through the shock wave near the $ z $-axis again.  
In model A4, the specific total energy increases
from $ E_{\rm tot} $ = $ -0.002 c^2 $ to $ 0.03 c^2 $ at the 
first passage and to $ 0.06 c^2 $ at the second passage.
The energy gain is due to the variability of the flow.
The specific total enthalpy ($ H + \mbox{\boldmath $ v $}^2/2 $) 
conserves in the shock rest frame.
It increases in the frame in which the shock wave travels.
The energy gain, $ \Delta E $, is proportional to the shock propagation speed 
and to the velocity difference at the shock,
\begin{equation}
 \Delta E =
 \mbox{\boldmath$ u $} \cdot
 ( \mbox{\boldmath$ v $}_2 - \mbox{\boldmath$ v $}_1 ) \; ,
\label{energy-gain}
\end{equation}
where $ \mbox{\boldmath$ u $} $ denotes 
the phase velocity of the shock front, and
the suffix \lq\lq 1'' and \lq\lq 2'' denote 
the value in pre- and post-shocked flow.
When the shock wave propagates to the upstream (downstream)
side, the gas passing through the shock wave gains (looses)
energy.

     The energy gain is larger when the shock wave is stronger
and moves faster.  The shock strength and the propagation speed
vary from model to model depending mainly on $ \ell $ of a gas torus.
When $ \ell < \ell_{\rm mb} = 2 c r_g $, 
the gas torus is absorbed into
the black hole without bounce and hence without forming
a shock wave.  When $ \ell $ is only a little larger
than $ \ell_{\rm mb} $, the gas torus bounces but only
weakly.  Hence the shock wave is weak and moves slowly.
When $ \ell \simeq 2.2 c r_g $, the bounce is
strongest. The resultant shock wave is strongest and
moves fastest.  When $ \ell > 2.2 c r_g $,
the bounce is weaker for a larger $ \ell $.
This dependence on $ \ell $ can be understood easily if 
we approximate the gas torus as a test particle moving
in the effective potential. When $ \ell = 2.2 
c r_g $, the effective potential has the deepest local minimum.

     In the following we consider the efficiency of 
acceleration from the opposite side.  When the jets are accelerated
and gain energy, the shock producing gas, i.e., the infalling
gas torus looses its energy. 
The energy gain and loss should balance.  The energy loss
has an upper bound given by the energy difference between
the initial and final states.  When the gas torus settles at
the bottom of the effective potential, it liberates maximum
energy available.  In model A4 the specific total energy of the gas torus is
$ - 0.022 c^2 $ at the bottom of the effective
potential in the final state.
Hence the gas torus can liberate 
at most $ \Delta E = 0.02 c^2 $ per unit mass. 
This means that a gas torus of $ 6 \times 10^{28} $ g
can liberate the total energy of $ 1.2 \times 10^{48} $ erg.  
We measure the efficiency of jet formation, $ \eta $, by the ratio of
the total energy of the jets to the total available energy.
The efficiency is 3\% in model A4.  The efficiency is
listed for each model in Table 1.

\subsection{Comparison with Other Hydrodynamical Models}

In this subsection we compare our numerical simulations with others.
A large number of numerical simulations of jets have
been published thus far.  Some invoke magnetic fields 
(e.g., Blandford \& Payne 1982; Uchida \& Shibata 1985) and
some others do high luminosity as jet driving sources 
(e.g., Eggum et al.\ 1988).
Some study formation and propagation of jets 
(e.g., Hawley et al.\ 1984a, 1984b; Clarke et al.\ 1985; 
Molteni et al.\ 1994; Yokosawa 1994),
while some others do only propagation of jets 
(e.g., Tenorio-Tagle, Cant\'o, \& Ro\'zyczka 1988).  In this 
subsection we restrict ourselves to hydrodynamical
simulations in which jets emerge from accretion onto a
black hole.
Comparison between MHD models and ours will be discussed 
in the next subsection.

Hawley et al.\ (1984b) noticed outflow in their simulations
of accretion onto a black hole.  In contrast with our model,
the outflow is hollow and along the outer
edge of the thick accreting disk.  The gas density is
very low near the $ z $-axis, since the 
accreting gas has large angular momentum and
accordingly the centrifugal force is very strong near the $ z $-axis.

Clarke et al.\ (1985) studied more spherical accretion
onto a black hole. They assumed constant density and angular
velocity at the outer boundary set at a given radial distance
from the center.  Their simulations showed shock waves 
similar to those found in our models. 
$ \lbrack $A similar shock wave is found by Wilson (1972) 
although he did not find outflows.$ \rbrack $  
Also in their simulations the shock waves intersect themselves 
on the $ z $-axis and form \lq \lq density knots '' propagating outward.  
The  density knots are similar to the jets in our models although they
are too weak to evolve into jets.  Since the outer boundary
condition is fixed, the accretion approaches to a steady state
in which the central black hole absorbs almost all the 
accreting gas.

Molteni et al.\ (1994) studied accretion onto
a black hole using the smoothed particle hydrodynamics (SPH).
Although their methods are different, their accretion flow is similar
to that of Hawley et al.\ (1984b).  The accreting gas has
large angular momentum and the density is very low near the $ z $-axis.
Although the mass ejection rate is high in their model, 
it is mainly because the accreting gas has high energy and 
can be ejected without energy gain.

Yokosawa (1994) studied accretion taking account of viscosity.
Although the gas is concentrated near the midplane at his initial
model, the polar region is filled with low specific angular momentum
gas at later stages.  The viscosity transfers the angular
momentum and produces low specific angular momentum gas.  
The low specific angular momentum gas is ejected 
as polar (filled) jets also in Yokosawa (1994) although
the shock waves and jet acceleration 
are not clearly seen in his simulations.

It has been confirmed from the above comparisons that
(1) the polar jets consist of low specific angular momentum
gas and (2) the time variability is essential to
produce strong jets.  

\subsection{Comparison with MHD Models}

Simulations of jets from time-dependent MHD accretion flow have been performed 
by many authors (e.g., Uchida \& Shibata 1985; Shibata \& Uchida 1986;
Kudo \& Shibata 1997; Koide, Shibata, \& Kudoh 1998 and others).
Shibata \& Uchida (1986) investigated accretion disk
initially threaded by magnetic field lines, and
found the magnetically driven bipolar outflows.
The outflows transport angular momentum from the disk, and thus 
form hollow jets contrary to filled jets of our model.
In Shibata \& Uchida (1986) 
the mass ejected as the jets is about 10 \% of the original disk mass, and
the energy of the jets is about 10 \% of the potential energy released
by the accretion disk.
The mass ejection and the energy ejection rates of the MHD jets
is about twice as efficient as those in our typical jets.
Only model D1 has comparable ejection rates
to MHD models in our models.

In MHD models there is no strong shock waves 
during the jet acceleration processes,
while in our model strong shock waves form and accelerate jets.
This coexistence of shock waves and jets in our model 
gives a natural explanation of 
simultaneous X-ray and $ \gamma $-ray flares and 
the following relativistic jets ejections, 
which are observed in active AGNs and X-ray binaries,
as shown in the next subsection.

\subsection{Application to AGNs and Galactic Black Hole Candidates}

In this subsection we discuss the application of our model
to AGNs and Galactic black hole candidates, X-ray binaries.  
In our model highly variable accretion produces
transient shock waves, which accelerate jets.
Since the shock waves are strong and semi-relativistic,
they will produce high energy cosmic rays including $ \gamma $-rays.
The shock heated gas will emit thermal X-rays.  The $ \gamma $-rays
and thermal X-rays will be observed as flares since their
production is temporal.  Thus our model predicts simultaneous
flares in $ \gamma $- and X-rays and jets formation associated
with the flares.  

There are some observations indicating simultaneous
$\gamma $- and X-ray flares.  The BL Lac object, Mrk 421 flared up
in the X-rays on May 16, 1994.
Its doubling time scale was about 12 hr (Takahashi et al.\ 1996).
The $ \gamma $-ray flare preceded one day 
to the X-ray flare (Kerrick et al.\ 1995).

Positive correlation is found for other BL Lac objects 
between the flux and hardness of the X-ray (Sembay et al.\ 1993). 
When the source is brighter, the spectrum is harder, i.e.,
the non-thermal component is more dominant.  This correlation
supports an idea that transient shock waves cause
flares in the high energy range.

Although association of jets with flares has not been observed for
AGNs, it is observed for black hole candidates in our Galaxy.
Hjellming \& Rupen (1995) reported episodic ejection of
relativistic jets by the X-ray transient GRO J1655$-$40.
GRO J1655$-$40 flared up in the X-ray in August 1994.
In the VLA images the radio bright points seem to be ejected
from the black hole candidate with a semi-relativistic speed.
The epochs of the ejection coincide with the peaks of the
radio emission with probable delay of several days.  
This observation is consistent with our model.
Similar events are observed for the Galactic superluminal source 
GRS 1915+105 (Foster et al.\ 1996)

Our time-dependent model can produce observed semi-relativistic jets.
The velocity of jets observed in black hole candidates
ranges from sub-relativistic $ v_{\rm jet} \sim 0.26 c $ 
for the X-ray binaries, SS433 (Margon 1984), to ultra-relativistic 
for the BL Lac object Mrk421 (Gaidos et al.\ 1996).
In our model the velocity of the jets is typically $ 0.3 - 0.5 c $.
It is larger when the ambient flow is colder.
If we reduce the temperature of the ambient gas, we will obtain
faster jets.  It is, however, not likely to obtain ultra-relativistic
jets by our mechanism.  It is also true for other known mechanisms.
Ultra-relativistic jets would require other unknown acceleration mechanisms.

\section{CONCLUSIONS}

We investigated time-dependent accretion flows onto a massive 
black hole numerically.
We used new techniques in our simulations, the multi time step scheme for
time integration and the extended Roe (1981) type approximate Riemann solver
for general equation of state.
These techniques enabled us to perform numerical simulation
with high resolution and to follow long term evolution.
The high spatial resolution was essential to simulate a cold accretion flow.

We considered the accretion that consists of 
hot ambient gas with low specific angular momentum and
cold dense gas with high specific angular momentum.
The ambient gas accretes continuously and 
the dense gas highly intermittently as blobs.
Our main results are summarized as follows.

(1) Bipolar jets emanate when the infalling gas blobs have 
high specific angular momentum
$ ( \ell \geq \ell_{\rm mb} ) $.
The blobs form shock waves by colliding with the ambient gas
after the bounce at the centrifugal barrier.
A part of the shock heated gas outflows along the rotation axis and
evolves into pressure-driven jets.
The jet mass ejection is largest when the blobs with 
$ \ell = 2.2 c r_g $ infall.

(2) When gas blobs have low specific angular momentum
$ ( \ell < \ell_{\rm mb} ) $,
shock waves and jets are not formed.
The blobs accrete smoothly onto a black hole 
without feeling the centrifugal barrier.

(3) Jets are more massive when the blobs are more massive.
There is a critical mass of the blobs, $ \sim 10^{28} $ g, to emanate jets,
when the average accretion rate is $ 0.11 M_\odot $ yr$^{-1}$.

(4) The efficiency of the energy gain of the jets 
is $ 3 \sim 11 $\%.
It is larger when the blobs have higher specific angular momentum or 
larger mass.

(5) When the ambient gas is colder, the jets are better collimated.

Although we assumed strict axisymmetry in our model,
we expect that the jet acceleration mechanism 
will work also in quasi-axisymmetric accretion.
When the accretion is not strictly axisymmetric,
the shock front will be also not strictly axisymmetric.
The shock wave will converge but with offset and
the jet will be slightly offset from the $ z $-axis.
The non-axisymmetric accretion is one of the future
problem for three-dimensional numerical simulations.

\acknowledgments

KN acknowledges the support of the JSPS Research Fellowships 
for Young Scientists.
The numerical computations presented here were carried out mainly on
VPP300/16R and VX/4R at the Astronomical Data Analysis Center
of the National Astronomical Observatory, Japan. 
This work is supported in part by the Grant-in-Aid 
for the Priority Area (07247212).

\appendix
\section{Approximate Riemann Solver for a General Equation of State}

Our scheme is essentially based on the multi grid scheme
by Chiang et al.\ (1992) who solved
the flow of an ideal gas in the Cartesian coordinates.
They used the approximate Riemann solver of Roe (1981)
to evaluate the numerical flux.
We extended the approximate Riemann solver of Roe (1981) 
to include blackbody radiation.
Our Riemann solver can be applied to
any equation of state whenever
all the thermodynamical variables are 
expressed as a function of $ \rho $ and $ E $.
In the following we describe the numerical flux we used.

We solved the hydrodynamical equations 
(\ref{mass-con}) $ - $ (\ref{energy-con}) in the spherical polar 
coordinates assuming the axisymmetry.
The equations in the conservation form can be expressed,
\begin{equation}
{\partial \over \partial t} 
(\mbox{\boldmath$T$} \mbox{\boldmath$U$} )
+ {\partial \over \partial r}  
(\mbox{\boldmath$T$} \mbox{\boldmath$F$} )
+ {1 \over r} {\partial \over \partial \theta}
(\mbox{\boldmath$T$} \mbox{\boldmath$G$} ) = 
\mbox{\boldmath$S$} \; ,
\label{hydroeq}
\end{equation}
where
\begin{eqnarray} 
 &&
 \mbox{\boldmath $U$}=
 \left(
 \begin{array}{c}
 \rho \\
 \rho v_r \\
 \rho v_\theta \\
 \rho v_\phi \\
 \rho E
 \end{array}
 \right)
 \quad , \qquad
 \mbox{\boldmath $F$}=
 \left(
 \begin{array}{c}
 \rho v_r \\
 \rho v_r{}^2 +P \\
 \rho v_r v_\theta \\
 \rho v_r v_\phi \\
 \rho v_r H
 \end{array}
 \right)
\quad , \qquad 
 \mbox{\boldmath $G$}=
 \left(
 \begin{array}{c}
 \rho v_\theta \\
 \rho v_\theta v_r \\
 \rho v_\theta{}^2 +P \\
 \rho v_\theta v_\phi \\
 \rho v_\theta H
 \end{array}
 \right)
 \; ,
 \nonumber \\
  &&
  \mbox{\boldmath $T$}=
  \left(
  \begin{array}{ccccc}
  r^2 \sin \theta & 0 & 0 & 0 & 0 \\
  0 & r^2 \sin \theta & 0 & 0 & 0 \\
  0 & 0 & r^3 \sin \theta & 0 & 0 \\
  0 & 0 & 0 & r^3 \sin^2 \theta & 0 \\
  0 & 0 & 0 & 0 & r^2 \sin \theta 
  \end{array}
  \right) \; ,
  \nonumber \\
 &&
 \mbox{\boldmath $S$}=
 \left(
 \begin{array}{c}
 0 \\
 r \sin \theta ( \rho v_\theta^2 + \rho v_\phi^2 +2 P ) 
 + r^2 \sin \theta \rho g_r \\
 -r^2 \cos \theta ( \rho v_\phi^2 + P ) \\
 0 \\
 r^2 \sin \theta \rho v_r g_r
 \end{array}
 \right)
 \; ,\end{eqnarray}
and
\begin{equation}
 g_r = -{GM \over (r-r_g)^2}
 \; .
\end{equation}
The conservative variable vector, $ \mbox{\boldmath$U$} $,
is evaluated on the grid points, $ (r_i, \theta_j) $,
where $ i $ and $ j $ denote the grid numbers in the
$ r $- and the $ \theta $-directions, respectively.

The numerical flux between the grid points 
$ (r_i, \theta_j) $ and $ (r_{i+1}, \theta_j) $
is evaluated as 
\begin{equation}
\mbox{\boldmath $F$}_{i+1/2,j}^{n}
= 
{1 \over 2}
  \left(
  \mbox{\boldmath $F$}_{i,j}^{n}
  + \mbox{\boldmath $F$}_{i+1,j}^{n}
  - \sum_{k=1}^5 | \lambda_k | \delta w_k \mbox{\boldmath $R$}_k
  \right) \; ,
\end{equation}
where
\begin{eqnarray}
 \mbox{\boldmath $R$}_1
 =
 \left(
 \begin{array}{c}
 1 \\
 \bar v_r \\
 \bar v_\theta \\
 \bar v_\phi \\
 \bar v^2/2 + \varepsilon
 \end{array}
 \right)
 \; ,
&&
 \mbox{\boldmath $R$}_2
 =
 \left(
 \begin{array}{c}
 0 \\
 0 \\
 1 \\
 0 \\
 \bar v_\theta
 \end{array}
 \right)
 \; , \;
 \mbox{\boldmath $R$}_3
 =
 \left(
 \begin{array}{c}
 0 \\
 0 \\
 0 \\
 1 \\
 \bar v_\phi
 \end{array}
 \right)
 \; ,
 \nonumber
 \\
&& 
 \mbox{\boldmath $R$}_4
 =
 {\bar \rho \over 2 \bar c_s}
 \left(
 \begin{array}{c}
 1 \\
 \bar v_r + \bar c_s \\
 \bar v_\theta \\
 \bar v_\phi \\
 \bar H + \bar v_r \bar c_s
 \end{array}
 \right)
 \; , \;
 \mbox{\boldmath $R$}_5
 =
 {\bar \rho \over 2 \bar c_s}
 \left(
 \begin{array}{c}
 1 \\
 \bar v_r - \bar c_s \\
 \bar v_\theta \\
 \bar v_\phi \\
 \bar H - \bar v_r \bar c_s
 \end{array}
 \right) \; ,
\end{eqnarray}
\begin{equation}
\lambda_1 = \bar v_r \; , \;
\lambda_2 = \bar v_r \; , \;
\lambda_3 = \bar v_r \; , \;
\lambda_4 = \bar v_r + \bar c_s \; , \;
\lambda_5 = \bar v_r - \bar c_s \; ,
\end{equation}
\begin{equation}
\delta w_1 = \rho_{i+1,j} - \rho_{i,j} 
 - { P_{i+1,j} - P_{i,j} \over \bar c_s^2 } \; ,
\end{equation}
\begin{equation}
\delta w_2 = v_{\phi,i+1,j} - v_{\phi,i,j} \; ,
\end{equation}
\begin{equation}
\delta w_3 =  - v_{\theta,i+1,j} + v_{\theta,i,j} \; ,
\end{equation}
\begin{equation}
\delta w_4 = v_{r,i+1,j} - v_{r,i,j} 
+ { P_{i+1,j} - P_{i,j} \over \bar \rho \bar c_s } \; ,
\end{equation}
\begin{equation}
\delta w_5 = - v_{r,i+1,j} + v_{r,i,j} 
+ { P_{i+1,j} - P_{i,j} \over \bar \rho \bar c_s } \; ,
\end{equation}
\begin{equation}
\bar \rho = \sqrt{\rho_{i,j} \rho_{i+1,j}} \; ,
\end{equation}
\begin{equation}
\bar v_r = 
{ \sqrt{\rho_{i,j}} v_{r,i,j} + \sqrt{\rho_{i+1,j}} v_{r,i+1,j} 
  \over
  \sqrt{\rho_{i,j}} + \sqrt{\rho_{i+1,j}}
} \; ,
\end{equation}
\begin{equation}
\bar v_\theta = 
{ \sqrt{\rho_{i,j}} v_{\theta,i,j} + \sqrt{\rho_{i+1,j}} v_{\theta,i+1,j} 
  \over
  \sqrt{\rho_{i,j}} + \sqrt{\rho_{i+1,j}}
} \; ,
\end{equation}
\begin{equation}
\bar v_\phi = 
{ \sqrt{\rho_{i,j}} v_{\phi,i,j} + \sqrt{\rho_{i+1,j}} v_{\phi,i+1,j} 
  \over
  \sqrt{\rho_{i,j}} + \sqrt{\rho_{i+1,j}}
} \; ,
\end{equation}
\begin{equation}
\bar v^2 = \bar v_r^2 + \bar v_\theta^2 + \bar v_\phi^2 \; ,
\end{equation}
\begin{equation}
\bar H = 
{ \sqrt{\rho_{i,j}} H_{i,j} + \sqrt{\rho_{i+1,j}} H_{i+1,j} 
  \over
  \sqrt{\rho_{i,j}} + \sqrt{\rho_{i+1,j}}
} \; ,
\end{equation}
\begin{equation}
\bar c_s^2 = ( \gamma -1 ) ( \bar H - \bar v^2 / 2 ) \; ,
\end{equation}
\begin{equation}
{ 1 \over \gamma -1 } =
{ \sqrt{\rho_{i,j}} h_{i,j} + \sqrt{\rho_{i+1,j}} h_{i+1,j}
  \over
  \sqrt{\rho_{i,j}} (\partial P / \partial \rho )_{s,j} 
    + \sqrt{\rho_{i+1,j}} (\partial P / \partial \rho )_{s,i+1,j}
} \; ,
\end{equation}
and
\begin{equation}
\varepsilon 
= 
{ ( \rho_{i+1,j} E_{i+1,j} - \rho_{i,j} E_{i,j} )
    - ( P_{i+1,j} - P_{i,j}) / ( \gamma -1 )
  \over
  ( \rho_{i+1,j} - \rho_{i,j} ) - ( P_{i+1,j} - P_{i,j}) / \bar c_s^2
} .
\end{equation}
This numerical flux is based on the flux vector splitting.
The differences in $ \mbox{\boldmath$U$} $ and 
$ \mbox{\boldmath$F$} $ are expressed
as the linear combination of the simple waves,
\begin{equation}
 \mbox{\boldmath $U$}_{i+1,j} - \mbox{\boldmath $U$}_{i,j}
 =
 \sum_{k=1}^5 \delta w_k \mbox{\boldmath $R$}_k
\end{equation}
and
\begin{equation}
 \mbox{\boldmath $F$}_{i+1,j} - \mbox{\boldmath $F$}_{i,j}
 =
 \sum_{k=1}^5 \lambda_k \delta w_k \mbox{\boldmath $R$}_k
 \; .
\end{equation}
This numerical flux reduces to that of Roe (1981)
when $ \gamma $ is constant.  The numerical flux 
in the $ \theta $-direction is evaluated in the same way.

Our scheme is similar to that of Glaister (1988). 
The main difference is the choice of sound speed. 
He uses the average pressure derivative to derive the sound speed, i.e., 
\begin{equation}
  \bar c_s^2 = {\bar P \bar P_E \over \bar \rho^2} + \bar P_\rho \; ,
\end{equation}
where
\begin{equation}
  \bar P_E = 
  { 
    \left\{
      \displaystyle{1 \over 2} 
      \left[ P(\rho_{i+1,j},E_{i+1,j})+P(\rho_{i,j},E_{i+1,j}) \right]
      - \displaystyle{1 \over 2} 
      \left[ P(\rho_{i+1,j},E_{i,j})+P(\rho_{i,j},E_{i,j}) \right]
  \right\}
  \over E_{i+1,j}-E_{i,j} 
  }
\end{equation}
and 
\begin{equation}
  \bar P_\rho =
  {
    \left\{
      \displaystyle{1 \over 2} 
      \left[ P(\rho_{i+1,j},E_{i+1,j})+P(\rho_{i+1,j},E_{i,j}) \right]
      - \displaystyle{1 \over 2} 
      \left[ P(\rho_{i,j},E_{i+1,j})+P(\rho_{i,j},E_{i,j}) \right]
    \right\} 
  \over  \rho_{i+1,j}-\rho_{i,j} 
  }
  \; ,
\end{equation}
while we use the average sound speed, $ \bar c_s $.
Our scheme provides a better approximation to the sound speed,
especially when the gas transit from a phase to another between the 
two adjacent points. Note that $ \bar c_s $ is real in our scheme 
as far as $ (\partial P / \partial \rho )_s $ is positive on the grid points.
The sound speed derived from the average pressure derivative can be imaginary.
Shimizu (1995) adopted our approximate Riemann solver 
in his numerical simulation of core collapse of supernovae 
and succeeded in capturing shock waves without numerical oscillations.


\newpage

\begin{figure}
\caption{
Model A4 is shown by the cross sections 
in the $ (\varpi, z ) $ plane.
Panels (a) through (f) are arranged in the time sequence; 
(a) $ t = 0 $ hr, (b) $ t = 1.9 $ hr, (c) $ t = 3.1 $ hr, 
(d) $ t = 4.8 $ hr, (e) $ t = 9.8 $ hr, and (f) $ t = 18.5 $ hr.
The arrows denote the velocity in the plane.
The contours denote the density.
The contour levels are spaced with an interval of 
$ \Delta (\log_{10} \rho) = 0.5 $.
The number, $ 10^{-14} $, specifies the contour of 
$ \rho = 10^{-14} $ g cm$^{-3}$.
Note that each panel is magnified at a different level.
\label{a2z}}
\end{figure}

\begin{figure}
\caption{
(a) Specific angular momentum distribution is shown by colors and contours 
at $ t $ = 4.8 hr for model A4 .
Contour levels are from $ \ell $ = 0.8 to $ 2.0 c r_g $ 
with an interval of $ 0.2 $.
The jet gas has low specific angular momentum.
(b) Same as (a) but for the specific total energy distribution.
Contour levels are $ E_{\rm tot} $ = 0.00, 0.02, 0.04, and $ 0.06 c^2 $.
The jet gas has positive specific total energy.
\label{jet7-L-Etot}}
\end{figure}

\begin{figure}
\caption{
The structures of jet at $ t $ = 9.8 hr in model A4.
The head of the jet reaches $ z = 48 r_g $ at this stage. 
(a) The density, (b) the radial velocity, (c) the Mach number, and 
(d) the specific total energy.
In panel (d), the solid, dotted, and dashed curves represent 
$ E_{\rm tot} $, $ H + \Phi $, and $ \Phi $, respectively.
\label{jet7-rho-v}}
\end{figure}

\begin{figure}
\caption{
The jet mass as a function of the time, $ t $, 
for models A3, A4, A5, and A6.
The numbers attached to the curves denote 
the specific angular momentum of the gas tours.
Model numbers are shown in the parentheses.
The mass ejection is the largest in model A5 ($ \ell = 2.2 c r_g $).
\label{jet-mass-L1}}
\end{figure}

\begin{figure}
\caption{
(a) The same as Fig.\ref{jet-mass-L1} but for models A4, B1, and B2.
The mass ejection is larger when the gas torus is more massive.
The number attached to each curve specifies the torus mass in the unit of 
$ 10^{27} $ g for each model.
(b) The ratio of the jet mass to the gas tori mass is 
plotted as a function of $ t $ for models A4, B1, and B2.
The mass ejection rate is proportional to the torus mass.
\label{jet-mass-m}}
\end{figure}

\begin{figure}
\caption{
The same as Fig.\ref{a2z} but for model C4.
Panels (a) through (f) are arranged in the time sequence; 
(a) $ t = 0 $ hr, (b) $ t = 1.9 $ hr, (c) $ t = 3.1 $ hr, 
(d) $ t = 4.8 $ hr, (e) $ t = 9.8 $ hr, and (f) $ t = 16.5 $ hr.
\label{a2z-c4}}
\end{figure}

\begin{figure}
\caption{
The same as Fig.\ref{jet-mass-L1} but for two-gas-tori models 
C1, C2, C3, and C4.
The evolution of models C1 and C2 is almost same.
The mass ejection is larger when the gas tori have 
larger specific angular momentum.
\label{jet-mass-L}}
\end{figure}

\begin{figure}
\caption{
The same as Fig.\ref{jet-mass-L1} but for models C4, D1, D2, F1,and F2.
In models C4, D1, and D2, only the shape of the initial gas tori is different.
The difference of the jet mass among three models is 
within a factor of $ 1.5 $.
The difference between models F1 and F2 is the variability of the flow.
When the variability of the flow is more mild the mass ejection is smaller.
\label{jet-mass-spc}}
\end{figure}

\begin{figure}
\caption{
The same as Fig.\ref{a2z} but for model E2.
Panels (a) through (f) are arranged in the time sequence; 
(a) $ t = 0 $ hr, (b) $ t = 1.9 $ hr, (c) $ t = 3.1 $ hr, 
(d) $ t = 4.8 $ hr, (e) $ t = 7.3 $ hr, and (f) $ t = 10.3 $ hr.
Jets have a higher gas velocity and thus are collimated more 
in model E2 than in model C4.
\label{a2z-e2}}
\end{figure}

\begin{figure}
\caption{
The structures of jet at $ t $ = 10.3 hr in model E2.
The head of the jet reaches $ z = 102 r_g $ at this stage. 
(a) The density, (b) the radial velocity, (c) the Mach number, and 
(d) the specific total energy.
In panel (d) the solid, dotted, and dashed curves represent 
$ E_{\rm tot} $, $ H + \Phi $, and $ \Phi $, respectively.
\label{jet15-rho-v}}
\end{figure}

\begin{figure}
\caption{
The same as Fig.\ref{jet-mass-L1} but for models C2, C4, E1, and E2.
When the ambient gas is colder, the jet ejection and
the jet mass are larger.
\label{jet-mass-amb}}
\end{figure}

\begin{figure}
\caption{
The jet formation mechanism is shown schematically.
The colors denote the density distribution in model A4.
Panels (a), (b), and (c) are arranged in the time sequence: 
(a) $ t = 0 $ hr, (b) $ t = 2.9 $ hr, and (c) $ t= 6.0 $ hr.
The thick solid curves denote the shock waves.
The dashed curve denotes the centrifugal barrier 
for the specific angular momentum $ \ell = 2.1 c r_g $.
The arrows denote the gas flow.
The torus resides within the dotted curve.
The labels denote the form of energy.
The energy changes its form in the order shown by the numbers.
The abbreviations \lq \lq G.E.'' and \lq \lq K.E.''
mean \lq \lq gravitational energy'' and \lq \lq kinetic energy'', respectively.
\label{jet-acceleration}}
\end{figure}

\end{document}